\documentclass[prl,twocolumn,showpacs,epsfig,longeq]{revtex4}
\usepackage{graphicx}
\usepackage{dcolumn}
\usepackage{amssymb}
\usepackage{bm}
\usepackage{epsfig}
\usepackage{psfrag}
\RequirePackage[pdftex,pdfpagemode=none, pdftoolbar=true,
                pdffitwindow=true,pdfcenterwindow=true]{hyperref}

\def\3dots{\:\raisebox{-0.5ex}{$\stackrel{\textstyle.}{:}$}\:}
\def\beq{\begin{equation}}
\def\eeq{\end{equation}}
\def\bea{\begin{eqnarray}}
\def\eea{\end{eqnarray}}
\begin{document}

\title{Raman spectroscopy of graphene on different substrates and influence of defects}

\author{Anindya Das, Biswanath Chakraborty and A.K. Sood$^{*}$}

\affiliation{Department of Physics, Indian Institute of Science, Bangalore - 560 012, India}

\begin{abstract}
We show the evolution of Raman spectra with number of graphene layers on different substrates, SiO$_{2}$/Si and conducting indium tin oxide (ITO) plate. The G mode peak position and the intensity ratio of G and 2D bands depend on the preparation of sample for the same number of graphene layers. The 2D Raman band has characteristic line shapes in single and bilayer graphene, capturing the differences in their electronic structure. The defects have a significant influence on the G band peak position for the single layer graphene: the frequency shows a blue shift upto 12 cm$^{-1}$ depending on the intensity of the D Raman band, which is a marker of the defect density. Most surprisingly, Raman spectra of graphene on the conducting ITO plates show a lowering of the G mode frequency by $\sim$ 6 cm$^{-1}$ and the 2D band frequency by $\sim$ 20 cm$^{-1}$. This red-shift of the G and 2D bands is observed for the first time in single layer graphene.
\end{abstract}

\maketitle
{\textit{*Corresponding author. A. K. Sood; Department of Physics, Indian Institute of Science, Bangalore - 560 012, India; Email: asood@physics.iisc.ernet.in\\
Manuscript is based on the invited talk delivered at National Review and Coordination Meeting on Nanoscience and Nanotechnology, 2007, Hyderbad, India.}}

\section{Introduction}
	The recent discovery\cite{NovScience2004,NovNature2005,NovPnas2005,ZhangNature2005} of thermodynamically stable two-dimensional single and bi-layer graphene has brought out many exciting experimental and theoretical studies, including the recent discovery of quantum Hall effect at room temperature\cite{NovScience2007}. The electronic properties near the Brillouin zone are governed by the Dirac equation, leading to the rich physics of quantum electrodynamics\cite{katsnelson}. The ballistic transport and high mobility\cite{NovNature2005,ZhangNature2005} in graphene make it a potential candidate for future nano-electronic devices\cite{Lemme,kimribbon,avouris}. Furthermore, graphene is a building block for other carbon allotropes like nanotube and graphite and hence its study will help in understanding the nanotubes.					
	
	Raman spectroscopy is one of the most powerful characterisation technique for carbon materials -- be it three dimensional form like diamond, graphite, diamond like carbon and amorphous carbon; two-dimensional graphene; one dimensional carbon like nanotubes; and zero demensional carbon like fullerenes. Graphite has three most intense Raman features at $\sim$ 1580 cm$^{-1}$ (G band), $\sim$ 1350 cm$^{-1}$ (D-band) and $\sim$ 2700 cm$^{-1}$ (2D band). A weak band at $\sim$ 3248 cm$^{-1}$, called 2D$\acute{}$ band, is an overtone of D$\acute{}$ (1620 cm$^{-1}$) mode. The G band is due to doubly degenerate E$_{2g}$ mode at the Brillouin zone center, whereas D band arises from defect mediated zone-edge (near K-point) phonons. The 2D band originates from second order-double resonant Raman scattering from zone boundary K+$\Delta$K phonons. The intensity of the Raman 2D band can be finite even if the D-mode Raman intensity is neglisible. It has been recently shown that Raman scattering can be used as a finger print for single, bi and a few layer graphene\cite{FerrariPrl2006,GuptaNanoLett2006,GrafNanoLett2007}. Recent experiments also show that the phonon frequency of the G mode of graphene and longitudinal optical-G mode of metallic nanotube blue shifts significantly by both electron and hole doping acheived by electric field gating in a field effect transistor geometry\cite{SimoneNaturematerials2007,YanPrl2007,Das2007,anindya,LAzzeriPrl2006,srijan}. In this paper, we present our work on the evolution of Raman spectra of the G and 2D modes for different graphene layers on Si/SiO$_{2}$ substrate. We will show that the G band peak position and I(G)/I(2D) ratio can not be taken as a finger print to identify a single layer graphene. On the other hand, 2D Raman band shape and position are good fingerprints of single and bilayer graphene to distinguish them from the multilayers. We also study the behaviour of the G peak frequency as a function of defects present in the single layer graphene, characterised by the intensity of the D-band.
	
	Recently, there has been a debate about the minimal conductivity in graphene due to impurities present on the SiO$_{2}$ substrate\cite{KimCondmat,SarmaCondmat}. These impurities could be inhomogeneous charge puddles on the SiO$_{2}$ substrate. We will show that the signature of these impurities is manifested in a finite blue shift ($\sim$ 5-7 cm$^{-1}$) of the frequency of the G mode. It is, therefore, interesting to study the properties of graphene on different substrates. Here we report, for the first time, the Raman spectra of single and few layers graphene on the conducting indium tin oxide (ITO) plates. In contrast to SiO$_{2}$/Si substrate, the G mode frequency of the single and few-layer graphene on the ITO plate is red shifted ($\sim$ 6 cm$^{-1}$) as compared to that of bulk HOPG ($\sim$ 1580 cm$^{-1}$). A large softening ($\sim$ 20 cm$^{-1}$) is also observed in the 2D mode of the single and few-layer graphene on ITO as compared to the garphene on SiO$_{2}$ ($\sim$ 2682 cm$^{-1}$). Such a lowering of frequency of the G as well as 2D modes has not been reported so far.
	
\begin{figure}[tbp]
\begin{center}
\leavevmode
\includegraphics[width=0.4\textwidth]{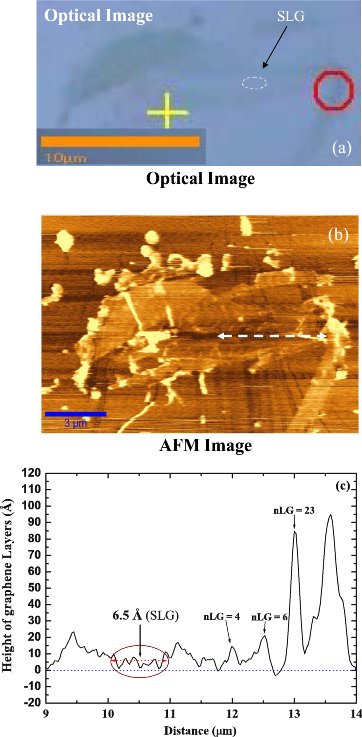}
\caption{(color online). (a) Optical image of a graphene flakes on 900 thick SiO$_{2}$/Si. The white marked region shows the single layer graphene (SLG). (b) AFM image of the above flake. (c) AFM line scan for height analysis along the white marked line in AFM image. SLG and nGL correspond to single and number of graphene layers, respectively.}
\label{Figure 1}
\end{center}
\end{figure}
	
\section{Experimental Details}			
	Graphene samples are prepared by micromechanical cleavage of bulk highly oriented pyrolitic graphite (HOPG) and are deposited on SiO$_{2}$/Si and transparent ITO plate. We have adopted two methods to make the graphene samples. In method A, a small piece of HOPG is glued to a glass slide using photoresist as an adhesive\cite{NovScience2004}. Then a scotch tape is used to make the graphite flake thinner and thinner till no thicker graphite flake is visible on the glass slide. The next step is to transfer the thin flakes of graphite, almost invisible, from glass slide onto the SiO$_{2}$/Si substrate by dissolving the photoresist in acetone. Method B involves gentle pressing of the HOPG flake, adhered to a scotch tape, directly on top of a SiO$_{2}$/Si substrate, which leaves some thin graphene layers on the substrate. We have used two different thickness of SiO$_{2}$ (900 and 300 nm) on Si and find that 300 nm thickness gives more optical contrast to identify single and bilayer graphenes\cite{Casiraghi1}. Among the two methods of sample preparation, we find that the method B can produce better quality samples. Raman spectrum are recorded using WITEC confocal (X100 objective) spectrometer with 600 lines/mm grating, 514.5 nm excitation at a very low laser power level (less than 1 mW) to avoid any heating effect.

\begin{figure}[tbp]
\begin{center}
\leavevmode
\includegraphics[width=0.4\textwidth]{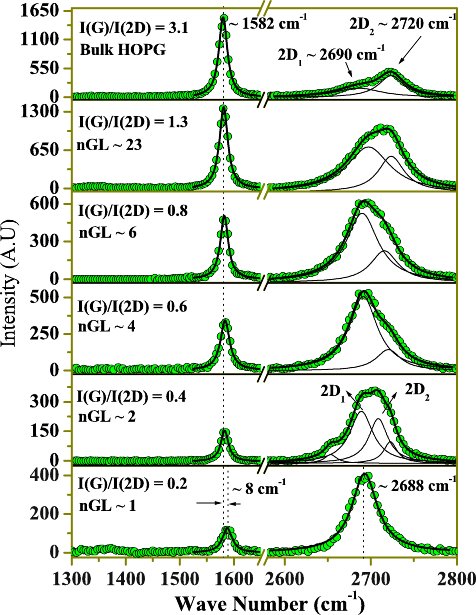}
\caption{(color online). Raman spectra of G and 2D mode for single, bi and few layers graphene. The open cirlces are the raw data and the solid lines are fitted Lorentzian functions.} 
\label{Figure 2}
\end{center}
\end{figure}

\section{Results and Discussions}
	We will divide the paper into three parts. In the first part we will discuss Raman signatures of different layers of graphene using method A of preparation on SiO$_{2}$/Si. In the second Part, we will see the influence of defects on Raman spectra of single layer graphene using method A and compare them with the spectra from the samples prepared using method B. In the last part we will report the Raman spectra of different graphene layers on a conducting ITO plate.
	
\subsection{Graphene on SiO$_{2}$/Si}
	Fig. 1a shows an optical image of graphene flake on 900 nm thick SiO$_{2}$/Si substrate. The marked portion indicates the region of single layer. The optical contrast is poor for the graphene on 900 nm thick SiO$_{2}$. To identify the single, bilayer and multilayer graphene, atomic force microscopy was used along with Raman spectroscopy. Fig. 1b shows an AFM image of the graphene  flake. The average height of the graphene layers is obtained from the AFM line scan analysis, as shown in Fig. 1c. In Fig. 1c we notice that the roughness of the SiO$_{2}$/Si substrate introduces noise in the height measurements. Due to different interactions of cantilever tip-substrate and cantilever tip-graphene, the height of single layer graphene is 0.6-0.7 nm for a single layer graphene, which is more than the van-der Waal separation between SiO$_{2}$ and graphene\cite{GuptaNanoLett2006}.

\begin{figure}[tbp]
\begin{center}
\leavevmode
\includegraphics[width=0.4\textwidth]{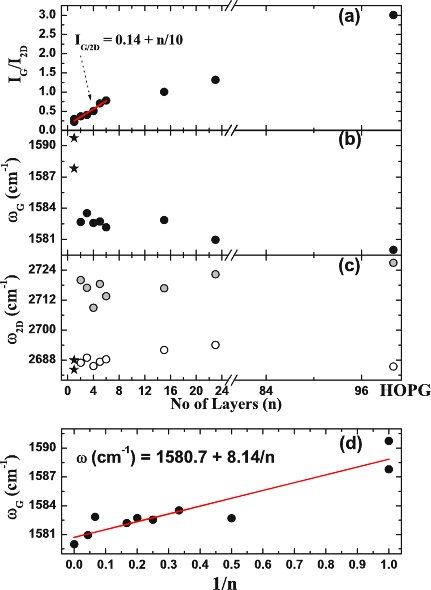}
\caption{(color online). (a) I(G)/I(2D) intensity ratio as a function of graphene layers. Peak position of (b) G band and (c) 2D band as a function of number of graphene layers (the stars correspond to the single 2D component of SLG). (d) G band peak frequency as a function of inverse of graphene layers.} 
\label{Figure 3}
\end{center}
\end{figure}	

	Fig. 2 shows Raman spectra (open circles) of different number of graphene layers, exhibiting G and 2D modes. Solid lines are the Lorentzian fit to the data. It is seen that the intensity ratio of G and 2D modes increases with the number of graphene layers. There is a blue shift of the G peak position ($\sim$ 8 cm$^{-1}$) in single layer graphene compared to the bulk HOPG. The shape of the 2D mode evolves significantly with the number of layers. The 2D mode in bulk HOPG can be decomposed in two components 2D$_{1}$ and 2D$_{2}$, whereas single layer graphene has a single component. The 2D Raman band in a bilayer is best fitted to four components. A single 2D component in monolayer and four components in bilayer graphene have been explained in terms of double resonance Raman scattering\cite{FerrariPrl2006}, which invokes electronic structure of the graphene layers. Therefore, the electronic structure of graphene is captured in its 2D Raman spectra. Another interesting observation is that in bulk HOPG, 2D$_{1}$ component is less intense than the 2D$_{2}$ component, whereas, for the bilayer they have almost the same intensity and furthermore increase of the number of layers leads to an increment of intensity of the higher frequency 2D$_{2}$ component compared to the 2D$_{1}$ component, as seen in Fig. 2. 
	
	In Fig. 3, we have plotted the I(G)/I(2D) intensity ratio, G peak position and 2D peak position as a function of numbers of graphene layers. The I(G)/I(2D) ratio increases almost linearly upto 6-8 layers, as shown in Fig. 3a. For a single layer, I(G)/I(2D) intensity ratio is $\sim$ 0.24, whereas for the bulk HOPG it is $\sim$ 3.2. The single layer graphene G mode appears at 1590 cm$^{-1}$, whereas for the bulk HOPG it is at 1580 cm$^{-1}$. Fig. 3b shows that the G peak position shifts to higher frequency as the number of graphene layers decreases and it varies almost linearly with the inverse of number of graphene layers, as shown in Fig. 3d (also shown in Ref\cite{GuptaNanoLett2006}). The single 2D peak of the monolayer graphene appears at 2685 cm$^{-1}$ which is red shifted with respect to the 2D$_{1}$ (2690 cm$^{-1}$) and 2D$_{2}$ (2720 cm$^{-1}$) peak positions of the bulk HOPG.

\begin{figure}[tbp]
\begin{center}
\leavevmode
\includegraphics[width=0.475\textwidth]{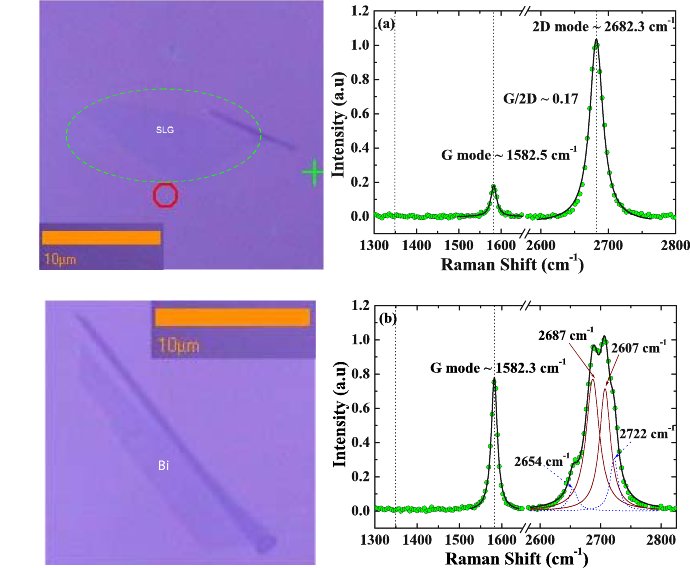}
\caption{(color online). Raman spectra of a (a) single layer graphene and (b) bilayer graphene on 300 nm SiO$_{2}$/Si using preparation method B. Optical images of single layer (top panel) and bilayer (bottom panel) graphene. Note that the D-band (expected near 1350 cm$^{-1}$) is absent for both single and bilayer graphene.}
\label{Figure 4}
\end{center}
\end{figure}
	
\subsection{Effects of defects}
	So far we have seen that the I(G)/I(2D) intensity ratio, G band peak position and the shape of the 2D band evolve with the number of graphene layers on the SiO$_{2}$/Si substrate using method A of preparation. Now we will look at the Raman spectra of the single layer graphene on SiO$_{2}$/Si using method B of preparation. Fig. 4 shows that there is no shift in G peak position ($\sim$ 1582 cm$^{-1}$) of the single layer and bilayer graphene with respect to the bulk HOPG. Therefore, G peak position can not uniquely identify the number of layers. It depends on how we prepare the samples. Recently, it has been shown that the G mode frequency increases as a function of both electron and hole doping\cite{SimoneNaturematerials2007,YanPrl2007,Das2007}. The blue-shifted G peak position observed in graphene prepared by method A can, therefore, be attributed to unintensional doping in graphene by the charge impurities present on the SiO$_{2}$/Si substrate.

\begin{figure}[tbp]
\begin{center}
\leavevmode
\includegraphics[width=0.475\textwidth]{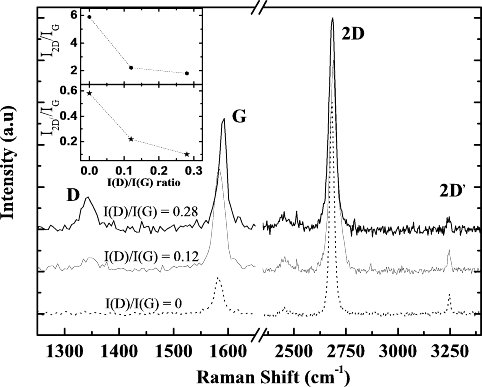}
\caption{(color online). Raman spectra of single layer graphenes for different I(D)/I(G) value. Inset shows dependence of I(2D)/I(G) and I(2D$\acute{}$)/I(G) on I(D)/I(G)}
\label{Figure 5}
\end{center}
\end{figure}	

\begin{figure}[tbp]
\begin{center}
\leavevmode
\includegraphics[width=0.4\textwidth]{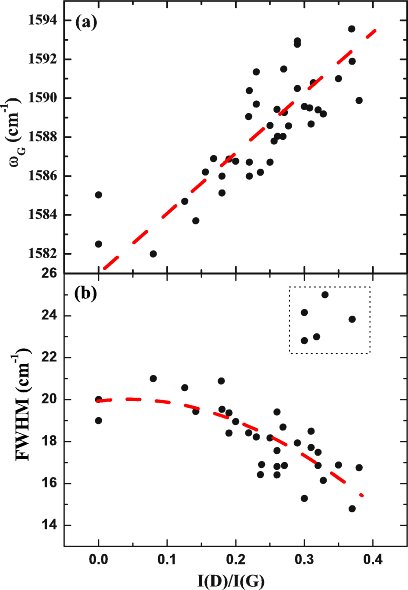}
\caption{(color online). (a) Peak position (b) FWHM of the G mode of different samples of single layer graphene as a function of I(D)/I(G) ratio. The dashed lines are guide to the eye.}
\label{Figure 6}
\end{center}
\end{figure}

	Another consequence of this unintensional doping is the presence of defects in single layer graphene. The defects could be edges, dislocations, cracks or vacancies in the sample. The so-called self-doping effects due to these defects is discussed in great detail by Neto \textit{et al.}\cite{Neto1,Neto2}. They have predicted that the Raman frequency of the G mode should be larger in presence of defects. We have recorded the Raman spectra of several single layer graphene flakes. The amount of defects present in the sample can be quantified by measuring the intensity ratio (I(D)/I(G)) of the D and G bands. Fig. 5 shows Raman spectra of single layer graphenes which have different amount of defects characterised by different I(D)/I(G) intensity ratio. We notice that the G band peak position increases with higher value of I(D)/I(G). We also note that there is a decrease in intensity ratio of the I$_{2D}$/I$_{G}$ and I$_{2D}$$\acute{}$/I$_{G}$ with more defects, as shown in inset of Fig. 5. Fig. 6 shows the G band peak position and FWHM as a function of I(D)/I(G) ratio for different single layer graphenes. The G frequency shift and decrement of FWHM are similar to the doping effects\cite{SimoneNaturematerials2007,YanPrl2007,Das2007}. We note that for some graphene flakes, the FWHM is higher ($\sim$24 cm$^{-1}$) with defects, as marked in a square box in Fig. 6b. This is opposite to the effect of doping, possibly due to the structural disorder as mentioned in ref\cite{Casiraghi2,Ferrari2001}. As predicted in ref\cite{Neto2}, we see that the G mode shifts almost linearly with defects. A maximum shift of the order of $\sim$ 10 cm$^{-1}$ is observed, which corresponds to a self doping of $\sim$ 10$^{13}$ cm$^{-2}$\cite{Das2007}.

\begin{figure}[tbp]
\begin{center}
\leavevmode
\includegraphics[width=0.475\textwidth]{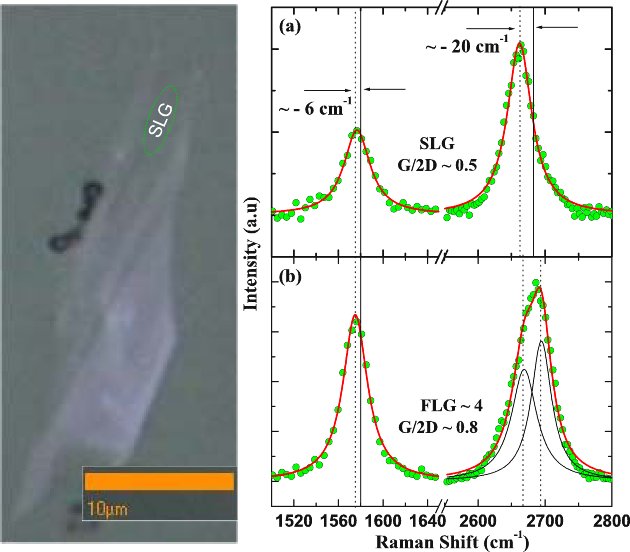}
\caption{(color online). Left panel shows the optical image of a graphene flake on ITO. Raman spectra of (a) single layer graphene and (b) few-layers graphene on ITO plate.}
\label{Figure 7}
\end{center}
\end{figure}

\begin{figure}[tbp]
\begin{center}
\leavevmode
\includegraphics[width=0.4\textwidth]{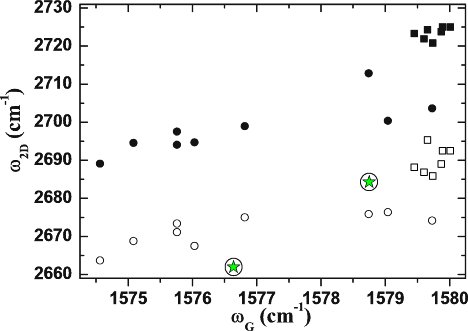}
\caption{(color online). 2D peak and G peak positions for single (stars inside a ring), a few layer (circles) and multi layer (squares) graphene on the ITO plate.}
\label{Figure 7}
\end{center}
\end{figure}

\subsection{Graphene on conducting ITO plate}
	As we have discussed above, the charge impurities on SiO$_{2}$/Si substrate and defects play a major role in determining the position of the G band of a single layer graphene. Therefore, it is interesting to look at the Raman spectra of graphene flakes on a different substrate like conducting indium tin oxide coated glass substrate. The graphene flakes were transferred onto a conducting ITO substrate using preparation method A. The left panel of Fig. 7 shows an optical image of graphene flake on the ITO plate. The marked region shows a single layer portion, as identified by the 2D Raman band. Optically it is very difficult to identify a single layer on the ITO plate, whereas we can see a colour contrast change for a few layer graphene. Fig. 7a and b show the Raman spectra of a single layer and four layer graphene on the ITO plate, respectively. The G and 2D bands appear at 1576 cm$^{-1}$ and 2665 cm$^{-1}$, respectively. It is interesting to note that the position of G and 2D bands are \textit{red shifted by $\sim$ 6 cm$^{-1}$ and $\sim$ 20 cm$^{-1}$}, respectively, compared to an undoped single layer graphene on the SiO$_{2}$/Si substrate. There is a correlation between the position of G and 2D modes as shown in Fig. 8 which plots the position of G and 2D mode of single layer (shown by stars inside a ring) and 2D$_{1}$ and 2D$_{2}$ modes of a few and multi layers graphenes. This is, to our knowledge, the first report of the red-shift of the G and 2D modes of the single layer graphene as compared to the corresponding Raman spectra on SiO$_{2}$/Si. The origin of the red-shift is not yet fully understood. The lowering of the frequency implies that the unit cell constant of the graphene layer is enlarged when deposited on the conducting ITO substrate. 
More studies are needed to understand the intriguing behaviour (red shift) of Raman specra of graphenes on conducting ITO plate.      
	
\textbf{Acknowledgement:}
AKS thanks Department of Science and Technology for funding the DST Unit in Nanoscience in IISc.

\end{document}